\documentclass[%
 reprint,
superscriptaddress,
 amsmath,amssymb,
 aps,
]{revtex4-2}

\usepackage{graphicx}
\usepackage{dcolumn}
\usepackage{bm}
\usepackage[colorlinks=true, allcolors=blue]{hyperref}
\usepackage[normalem]{ulem}
\usepackage{color}

\begin{document}


\title{Impulsive excitation of a solid by extreme contrast, high intensity femtosecond laser pulses}

\author{Sagar Dam}
\email{sagar.dam@tifr.res.in}
\affiliation{Tata Institute of Fundamental Research, 1 Homi Bhabha Road, Colaba, Mumbai 400005, India}

\author{J. F. Ong}%
\email{jianfuh.ong@eli-np.ro}
\affiliation{Extreme Light Infrastructure - Nuclear Physics (ELI-NP), ``Horia Hulubei'' National Institute for R\&D in Physics and Nuclear Engineering (IFIN-HH), 30 Reactorului Street, Bucharest-M\u{a}gurele, 077125, Romania
}

\author{Sk Rakeeb}
\affiliation{Tata Institute of Fundamental Research, 1 Homi Bhabha Road, Colaba, Mumbai 400005, India}

\author{Ameya Parab}
\affiliation{Tata Institute of Fundamental Research, 1 Homi Bhabha Road, Colaba, Mumbai 400005, India}

\author{Aparajit C}
\affiliation{Colorado State University, Fort Collins, CO 80523, USA}

\author{Anandam}
\affiliation{Tata Institute of Fundamental Research, 1 Homi Bhabha Road, Colaba, Mumbai 400005, India}

\author{Amit. D. Lad}
\affiliation{Tata Institute of Fundamental Research, 1 Homi Bhabha Road, Colaba, Mumbai 400005, India}

\author{Yash. M. Ved}
\affiliation{Tata Institute of Fundamental Research, 1 Homi Bhabha Road, Colaba, Mumbai 400005, India}

\author{M. Krishnamurthy}
\affiliation{Tata Institute of Fundamental Research, 1 Homi Bhabha Road, Colaba, Mumbai 400005, India}
\affiliation{Tata Institute of Fundamental Research Hyderabad, 36/P, Gopanpally Village, Serilingampally Mandal, Hyderabad, Telangana 500046, India}

\author{K. A. Tanaka}
\affiliation{Extreme Light Infrastructure - Nuclear Physics (ELI-NP), ``Horia Hulubei'' National Institute for R\&D in Physics and Nuclear Engineering (IFIN-HH), 30 Reactorului Street, Bucharest-M\u{a}gurele, 077125, Romania
}
\affiliation{University of Osaka, Yamadaoka, Suita, Osaka 565-0871, Japan}

\author{G. Ravindra Kumar}
\email{grk@tifr.res.in}
\affiliation{Tata Institute of Fundamental Research, 1 Homi Bhabha Road, Colaba, Mumbai 400005, India}

\date{\today} 

\begin{abstract}
We present the ultra-fast dynamics of the interaction between a high-intensity extreme contrast (expected to be around $\sim 10^{-18}$ at hundreds of picoseconds timescale) femtosecond laser and a solid. Simultaneous measurements of probe Doppler spectrometry and reflectivity in pump-probe experiments reveal the presence of extreme pressure in the solid density region, which triggers a long-lived ($\sim 15$ ps) strong inward shock. Hydrodynamic simulations accurately replicate these observations, providing a detailed explanation of the underlying physics.
\end{abstract}

\maketitle



Studies of ultrahigh-intensity, ultrashort laser interactions with solid targets offer exciting fundamental science and promise applications in particle acceleration, novel hard X-ray sources, nuclear fusion, and laboratory astrophysics \cite{fortov_Extreme_states_of_matter,Kruer,gibbonbook,proton_accelerator_snavely2000intense,proton_accelerator_robson2007scaling,electron_accelerator_kurz2021demonstration,electron_accelerator_hogan2005multi,electron_accelerator_litos2014high,electron_accelerator_rosenzweig1987nonlinear,ICF_atzeni2004beam,ICF_betti2016inertial,ICF_keefe1982inertial,ICF_myatt2014multiple}. The extreme rapidity of such interactions poses a challenge for the spatio-temporal study of the interaction region. Experimental probing of highly excited matter has utilized various techniques, using the pump-probe method in plasma reflectometry and Doppler spectrometry \cite{Amitavaadak_controlling_femto_shock,jana_ultrafast_dynamics,jana_reverse_shock,prashant_multicolor_probe,prashant_nature_shock,Amitavaultrafastdynamics}, polarimetry \cite{segre1997review,chatterjee2017micron}, and spectral interferometry \cite {ankit2022spectralinterferometry,ankit2024spectralinterferometryoptica}. These techniques have been employed from the femtosecond scale to tens of picoseconds, concomitant with particle-in-cell and hydrodynamic simulations. These studies have revealed shock waves induced by the intense pressure ($>10^{12}$ Pa) generated during the interaction \cite{jana_reverse_shock}. 

A major concern in the experiments has been the intensity contrast, which refers to the intensity in the low intensity wings of the amplified laser pulses in relation to the peak intensity. For a decade or so, high-intensity lasers have provided contrast levels of 10$^{-10}$. More recently, the use of plasma mirrors and pulse cleaning techniques has improved the contrast by another two to three orders of magnitude \cite{contrast_control1998mourou,contrast1998itatani,contrast_2002,plasmamirror2006wittmanntowards,plasmamirror2018foldes}.

It is essential to further improve the contrast, as the presently achievable and projected peak intensities are reaching highly super-relativistic levels.   Second-harmonic generation (SHG) is a straightforward and efficient method to achieve the square of the contrast of the fundamental pulse (subject to practical limitations in implementation and performance), as suggested long ago \cite{mourou2010temporal} and utilized in recent studies \cite{purvis2013relativistic,rocca2024ultra,bargsten2017energy,aparajit2021OL}. However, it has been discovered that the pulse width of a second-harmonic beam strongly depends on the input energy on the SHG crystal and reduces as the energy of the fundamental wavelength increases \cite{ref2}. In our case, the pulse duration is in the 100-200 femtosecond range, depending on the intensity. 

The interaction of extreme contrast pulses with solids,  that is,  the impulsive excitation of a high-density surface, is a highly intriguing phenomenon and key fundamental issues surrounding this interaction still remain unknown and unexplored. In this Letter, we delve into the dynamics of this interaction at femtosecond resolution using pump-probe reflectometry and Doppler spectrometry. We track the plasma dynamics to tens of picoseconds, demonstrating strong inward motion of the plasma density. We also observe an intriguing anticorrelation between the time variation of plasma reflectivity and the Doppler shift. Hydrodynamic simulations explain the observations and offer insights into the dynamics.

\begin{figure}[ht]
  \centering
  \includegraphics[width=3.4in]{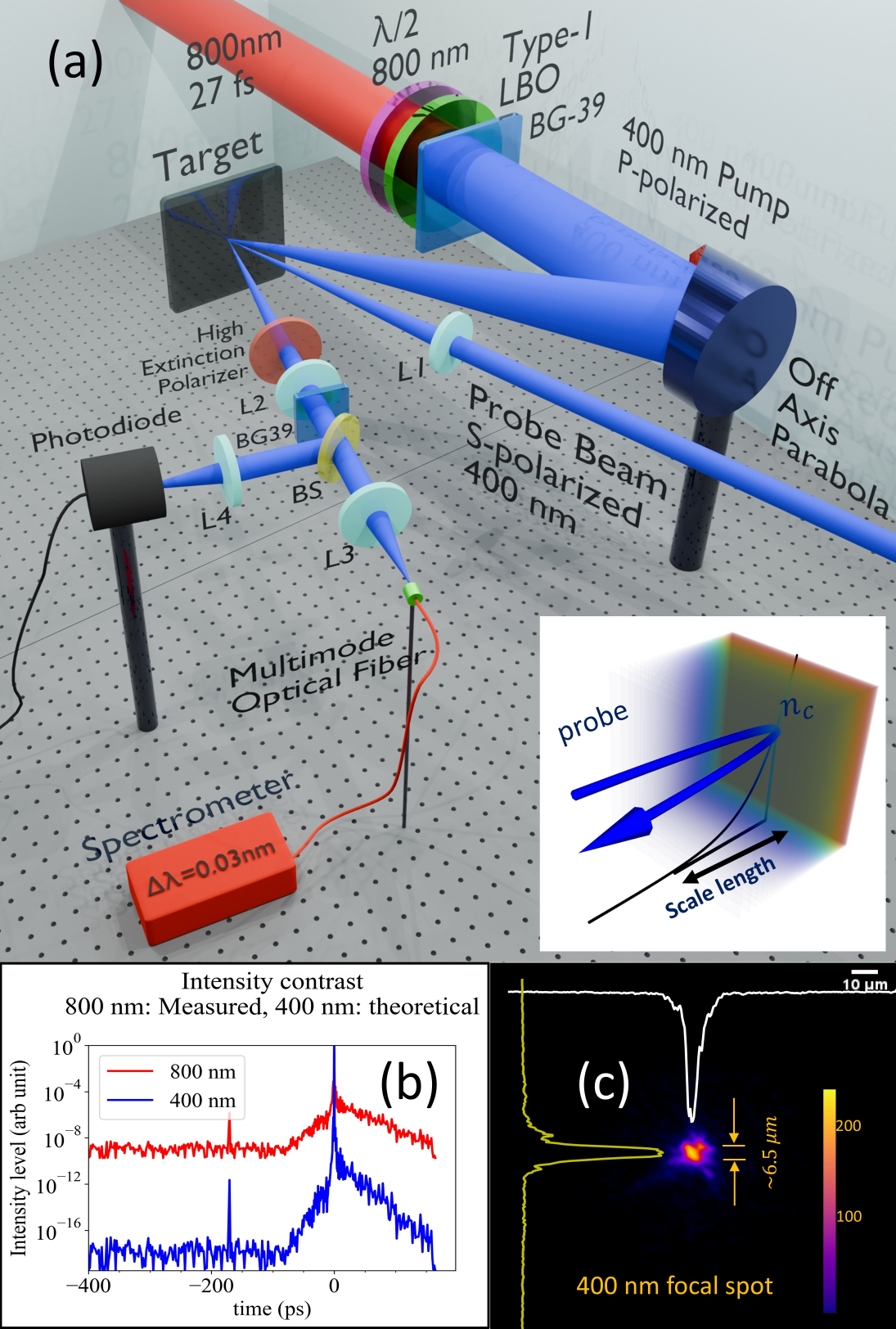}
  \caption{\label{fig:setup} (a) The Experimental setup schematic. In the actual setup, there are three $400 \, \mathrm{nm}$ dichroic mirrors used (along with the BG-39 filter) after the LBO to remove unconverted $800 \, \mathrm{nm}$. A high extinction polarizer (Nicol prism) is placed before the diagnostics to pass only the reflected probe. (b) Measured contrast profile for the $800 \, \mathrm{nm}$ fundamental beam and the theoretical contrast of $400 \, \mathrm{nm}$ SHG beam; (c) The $400 \, \mathrm{nm}$ focal spot at FWHM of $\sim 6.5 \ \mathrm{\mu m}$) on target.
  }
\end{figure}

The Tata Institute of Fundamental Research in Mumbai houses a Ti-Sapphire laser with a maximum peak power of 150 TW (Amplitude, France). This laser has a pulse width of $27 \, \mathrm{fs}$ and a central wavelength of $800 \, \mathrm{nm}$. It exhibits an intensity contrast of $10^{-9}$ at a $100 \, \mathrm{ps}$ timescale, measured using a third-order cross-correlation device (SEQUOIA, Amplitude). In the experiment depicted in Fig. \ref{fig:setup}, the laser pulse is split into a strong pump pulse and a weak probe pulse. The pump path utilizes an anti-reflection coated half-wave plate followed by a type-1 Lithium-tri-borate (LBO) for efficient conversion of the $800 \, \mathrm{nm}$ fundamental beam to a $400 \, \mathrm{nm}$ p-polarized second harmonic beam (SHG) \cite{aparajit2021OL}.

To optimize the conversion and polarization of the pump within the vacuum, the LBO is mounted on a dual angle control stage, which allows for adjustment of the incident angle $\theta$ and the transverse plane rotation angle $\phi$. The LBO output is reflected by three $400 \, \mathrm{nm}$ high-reflectivity mirrors, which efficiently attenuate the unconverted $800 \, \mathrm{nm}$ beam with an efficiency of $\frac{I_{800}^{\text{on target}}}{I_{800}^{\text{unconverted}}}<10^{-5}$. An additional thin BG-39 filter is employed to block the remaining unconverted $800 \, \mathrm{nm}$ beam. An off-axis $f$/3 dielectric parabolic focusing mirror (R $>$ 99\%) is used to focus the beam on a BK7 glass target (50 mm $\times$ 50 mm $\times$ 3 mm). A vacuum-compatible energy meter, mounted on a translational stage, can move in and out of the beam path for measuring the energy on target. The measured energy (for optimized SHG) exhibits a fluctuation of $\sim5\%$. The measured contrast of $800 \, \mathrm{nm}$ and the theoretically expected contrast of $400 \, \mathrm{nm}$ are illustrated in Fig. \ref{fig:setup} (b). The pump focal spot is measured to have a diameter of $\approx 6.5 \, \mathrm{\mu m}$ (Fig. \ref{fig:setup} (c)) at the target position.

The probe is converted to the second harmonic ($400 \, \mathrm{nm}$ s-polarized) using a 2 mm thick $\beta$-Barium-Borate (BBO) crystal, and reflected from a retro reflector mounted on a linear stage to control the delay between the pump and probe pulses. The unconverted 800-nm beam in the probe arm is blocked by a BG-39 bandpass filter. The probe is focused on the target using a 15 cm plano-convex lens. The spatial and temporal matching of the two pulses is verified using standard procedures \cite{jana_ultrafast_dynamics,jana_reverse_shock,Amitavaultrafastdynamics}. The details of the setup are given in the Supplementary Material (SM) \cite{SuppM}.

The probe reflects from the evolving critical surface or a high density hump on a lower density base of plasma for $400 \, \mathrm{nm}$ wavelength ($n_\text{cr}=\omega_L^2m_e\epsilon_0c^2/e^2 \approx 6.97 \times 10^{21} \, \text{cm}^{-3}$) \cite{Amitavaultrafastdynamics,Amitavaadak_controlling_femto_shock,sudiptamondal2010Doppler}. The probe is collected with an achromatic lens of focal length 15 cm, divided with a 200 $\mu$m thick fused silica beam splitter, and one part is sent to a photodiode (PD) connected to an oscilloscope and the other to a high-resolution triggerable spectrometer (HR-4000, resolution of $0.03$nm). A high extinction (extinction ratio $\sim 10^{-3}$) Nicol prism polarizer and a BG-39 filter are placed before the beam splitter to transmit only s-polarized beam and minimize noise from $800 \, \mathrm{nm}$ fundamental beam and $400 \, \mathrm{nm}$ LBO converted p-polarized pump beam. 

\begin{figure*}[htbp]
  \centering  \includegraphics[width=\textwidth]{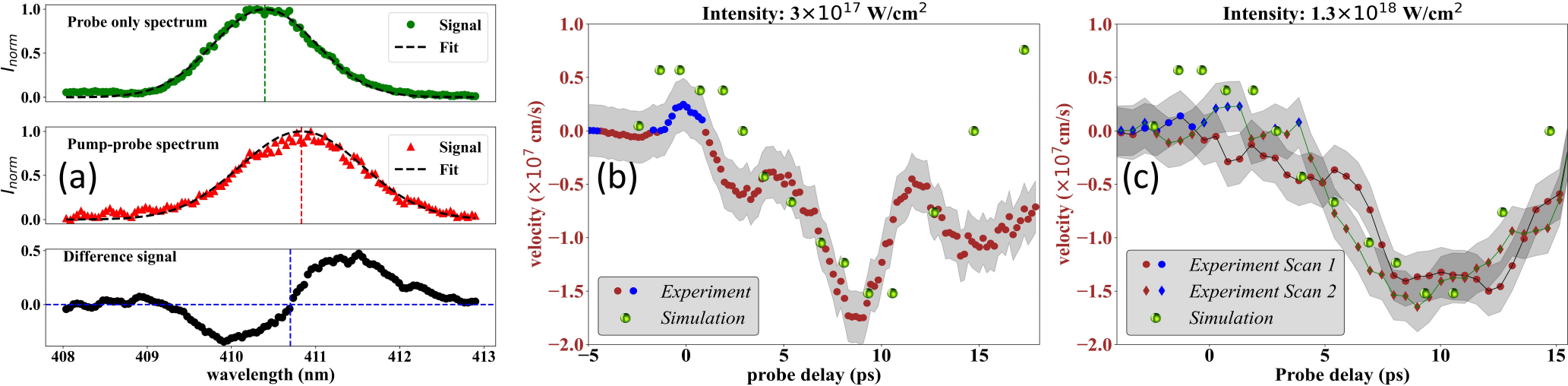}
  \caption{\label{fig:expdata}(a) The measured raw (normalized w.r.t peak) spectrum for probe only signal and pump-probe signal for some delay and the difference of pump-probe signal to probe only signal. The measured velocity of $400 \, \mathrm{nm}$ critical surface ($n_\mathrm{cr} \sim 6.97 \times 10^{21} \ \text{cm}^{-3}$ for $\lambda=$ $400 \, \mathrm{nm}$) has been shown for different intensities as (b)  $3 \times 10^{17} \, \mathrm{W \, cm^{-2}}$, (c) $1.3 \times 10^{18} \, \mathrm{W \, cm^{-2}}$.
  }
\end{figure*}

The reflected probe suffers a Doppler shift due to the motion of the pump excited surface. The shift is given by: \cite{Doppler_formula_moving_mirror}:

\begin{equation}
  \nu_\text{observed}= \nu_\text{source}\frac{1+\frac{2v}{c}\cos\alpha+\frac{v^2}{c^2}}{1-\frac{v^2}{c^2}} \label{eq:1}
\end{equation}

, where $\alpha$ is the incident angle of the probe on the target. As the probe incidence angle is very small, we can assume $\cos\alpha\approx1$ in Eq. ref{eq:1} and  approximate to
\begin{equation}
  v=-0.5c\frac{\Delta\lambda}{\lambda} 
\end{equation}
with $\Delta\lambda=(\lambda-\lambda_0)$ and $\lambda_0=c/\nu_\text{source}$\cite{ref6}. Here, the negative sign indicates that the plasma is moving into the target. We determine $\lambda_0$ as the central wavelength of the Gaussian-fitted input probe spectrum (averaged over 40-50 shots). For a given delay, $\lambda$ represents the central wavelength of the Gaussian-fitted reflected probe signal (Fig. \ref{fig:expdata} (a)). The standard deviation of the input probe signal is $\approx 0.064 \, \mathrm{nm}$, which is used to calculate the error bar for the pump-probe shots. 

Figures \ref{fig:expdata} (b) and (c) show the temporal evolution of the plasma density variation for intensities of $3\times10^{17} \, \mathrm{W \, cm^{-2}}$, and $1.3\times10^{18} \, \mathrm{W \, cm^{-2}}$, with pulse widths of 186 fs and 100 fs, respectively (as measured earlier in a frequency resolved optical gating (FROG) experiment \cite{ref2}). The red and blue colors indicate the critical surface moving inward (redshift) and outward (blueshift) from the target surface, respectively. The error bars represent the shot-to-shot fluctuations in the central wavelength, as indicated by the gray bands in the plots. A small outward motion can be observed for each intensity after the interaction with the main pulse, followed by a long inward motion. The reflected probe signal beyond 15 ps is weak for detection, however, the trend of reducing redshift is consistent with the simulation prediction, as discussed later.

\begin{figure}[htbp]
\centering
\includegraphics[width=3.4in]{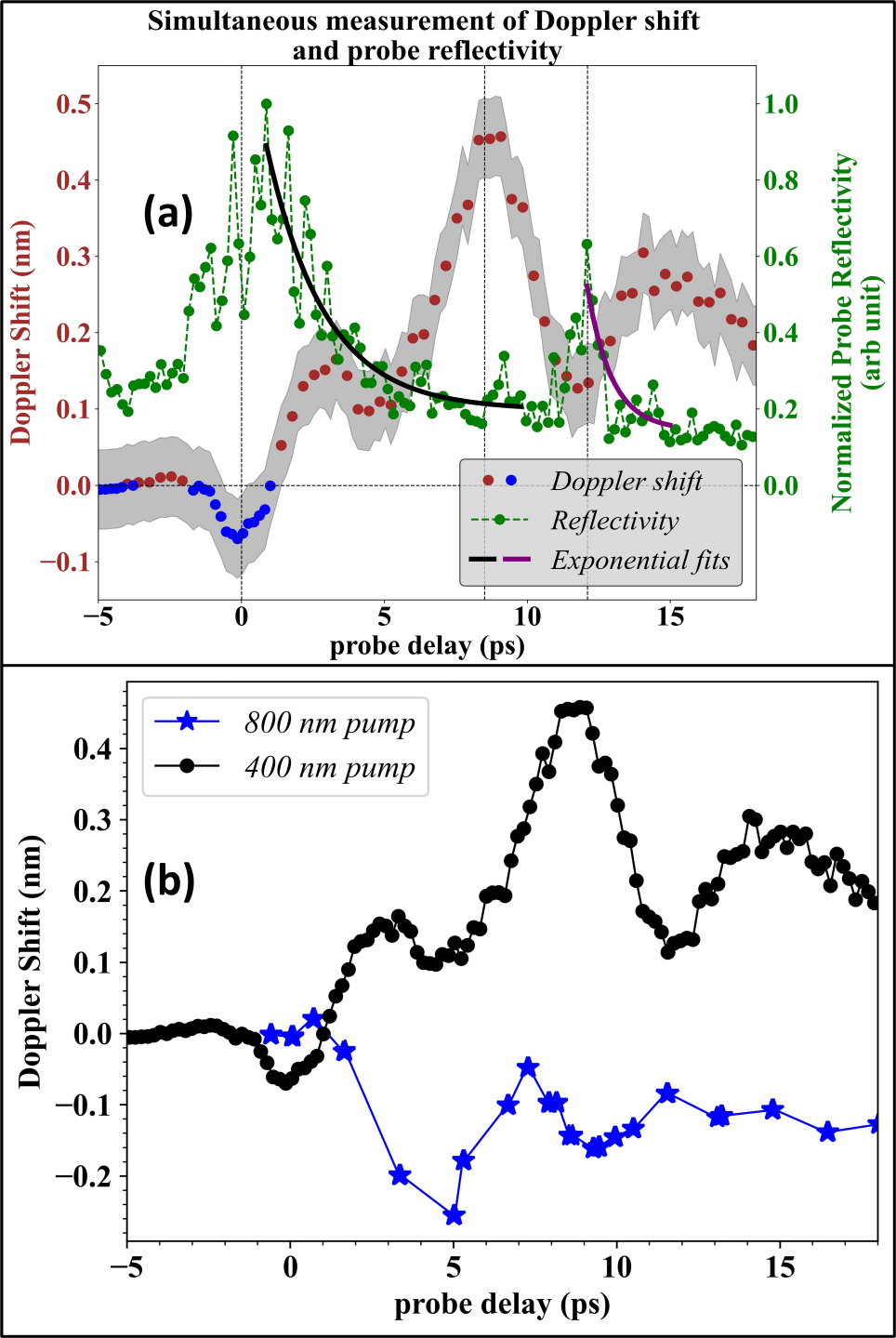}
  \caption{\label{fig:reflectivity}
  (a) Simultaneously measured probe reflectivity and Doppler shift for different pump-probe delays. The anti-correlations between reflectivity and Doppler shift are indicated by vertical dotted lines. The black and purple solid lines show the exponential fitting for reflectivity as $R_\text{normalized}(t)=e^{-\frac{t-t_\text{peak}}{\tau}}$, where $\tau$ is the time decay constant for reflectivity. In our case, we get $\tau=2.04$ ps for the first reflectivity peak (black fit) and 0.85 ps for the second peak (purple fit). (b) Comparison of Doppler shift for extreme contrast 400 nm current data vs moderately high contrast 800 nm earlier data (By Jana et al. \cite{jana_reverse_shock})}
\end{figure}

These experimental measurements show that the near-solid density matter receives an extreme impulse in a very short time due to the ultra-high contrast of the pulse that pushes matter inside for a long time. Earlier wavefront measurements at shorter timescales show a similar inward motion of the plasma critical surface \cite{rakeeb2025ponderomotive,rakeeb2025wavefront}.


We can also estimate the velocity of the critical surface by analyzing the simultaneously measured reflectivity of the plasma. As shown in Figure \ref{fig:reflectivity} (a), the reflectivity (at $I_\text{peak}\approx3\times10^{17}$ W cm$^{-2}$) increases when the pump interacts with the solid and creates plasma. However, at later delays, the plasma expands significantly with increased energy absorption, leading to a reduction in reflectivity. The drop in reflectivity can be fitted with an exponential curve: $R_\text{norm}(t)=e^{-\frac{t-t_\text{peak}}{\tau}}$, where $t_\text{peak}$ is the delay at which plasma reflectivity reaches its maximum.  The exponential decay time scale is measured to be $\tau=2.04$ ps.

The plasma absorption depends on the scale length of the underdense preplasma, which in turn depends on the velocity of the critical surface, i.e. $L(t)\sim\langle v_\text{crit}(t')\rangle_{t'}\times t$, where $\langle v_\text{crit}(t')\rangle_{t'}$ is the time averaged velocity for the critical surface from the time of peak reflectivity up to time delay $t$. Assuming the plasma density has a profile of $n_e(x)\sim n_\mathrm{cr}e^{-x/L}$, we can estimate the velocity from the relation of normalized reflectivity to the plasma scale length as \cite{Elizer,Kruer}: 

\begin{equation}
  R_\text{norm}(t)=\exp\left(-\frac{8\nu_{ei}L(t)}{3c}\right)
  \label{Eq:Rnorm}
\end{equation}

with the electron-ion collision frequency:

$$\nu_{ei}(T_e,n_i) \approx 2.9\times10^{-6}\ln\Lambda\frac{\overline Z^2n_i\ (\text{in cm}^{-3})}{[T_e\ (\text{in eV})]^{3/2}} \ \text{s}^{-1}.$$

where $n_i=n_e/\overline{Z}$ is the average ion density, $\overline{Z}$ is the average ionization state over different species inside the plasma, $T_e$ is the plasma electron temperature, and $\ln\Lambda$ is the Coulomb logarithm ($\Lambda= b_\text{max}/b_\text{min}$ where $b_\text{max}$ and $b_\text{min}$ are of the order of plasma Debye length and closest approach of free electron to ion). By taking $\overline{Z}\approx5$, $T_e\approx 135$ eV, and $\ln\Lambda\sim 10$ \cite{Plasma_formulatory}, we obtain the average velocity of the critical surface of $\sim 8.55\times10^6$ cm s$^{-1}$. Meanwhile, the average velocity up to 10 ps, calculated from the Doppler spectrometry measurement for $I_0 = 3\times10^{17}$ W cm$^{-2}$ (Fig. \ref{fig:expdata} (b)) is $7.43\times10^6$ cm s$^{-1}$. The two different approaches agree very well.

\begin{figure*}[htpp]
  \centering
  \includegraphics[width=\textwidth]{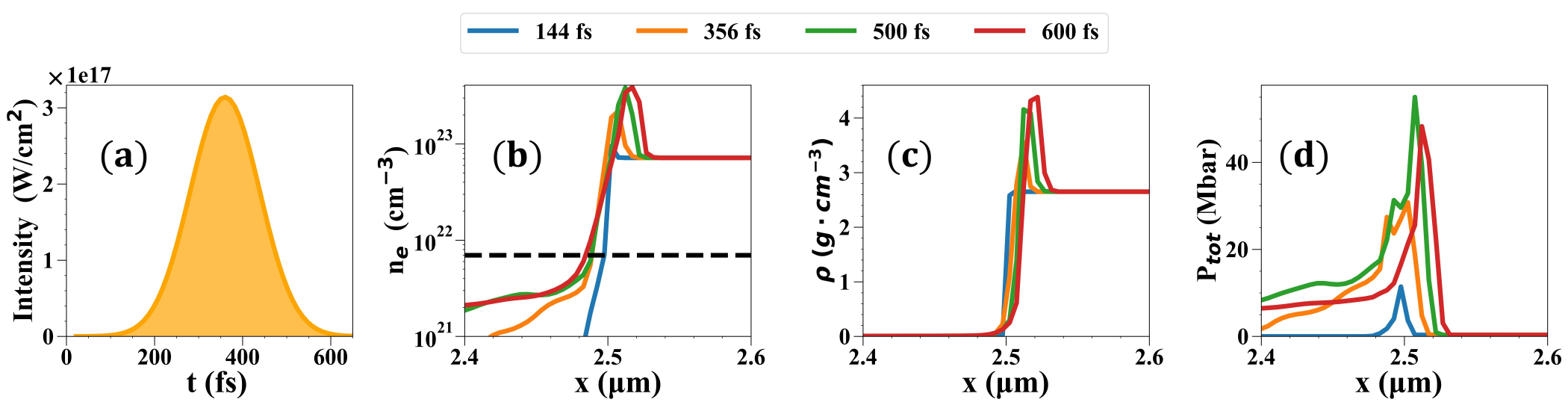}
  \caption{(a) The laser pulse profile for $I_0 = 3 \times 10^{17} \, \mathrm{W \, cm^{-2}}$ and FWHM pulse duration of $186 \, \mathrm{fs}$. The profiles of (b) the electron number density, (c) mass density, and (d) total pressure at different times of the corresponding laser pulse in (a). The dashed line in (a) indicates the probe critical density surface.}
  \label{fig:Y}
\end{figure*}

In addition, we also compare the present study performed with the 400 nm SHG pump beam, with the fundamental 800 nm pump beam (contrast $\sim 10^{-9}$ at 100 ps) \cite{jana_reverse_shock} in Fig. \ref{fig:reflectivity} (b). The motion of the critical density front is completely opposite for the 800 nm pulses. This implies that the prepulse of the fundamental pump with a lower contrast has caused a significant plasma expansion towards outer direction (showing strong reverse shock), and the inward compression wave could not be detected experimentally. There is however, an interesting correlation in the oscillatory period in both cases during plasma motion after the pump pulse leaves.

To understand the observations and get further insights into the dynamics, we performed one-dimensional (1-D) hydrodynamic simulations \textsc{flash} code version 4.8 \cite{Fryxell_2000}. We use the three-temperature treatment for plasma, tabulated equation of state (EoS), flux-limited diffusion model for electron conduction, heat exchange, multi-group opacity and radiation transport (see SM for simulation details \cite{SuppM}). Note that the FWHM pulse duration of $186 \, \mathrm{fs}$ is assumed for $I_0 = 3 \times 10^{17} \, \mathrm{W \, cm^{-2}}$.

Due to the extreme contrast of the second-harmonic laser, the laser pedestal at $t \leq -5 \, \mathrm{ps}$ does not result in significant preplasma generation. Therefore, we only model the interaction of the main pulse with a steep density gradient target. The laser energy deposition proceeds via inverse bremsstrahlung, while the resonance absorption ceases to be effective for steep density gradients \cite{PhysRevA.11.679}. In this situation, vacuum heating dominates and the absorption is $\geq  80 \%$ for $I_0 \lambda^2 \geq 10^{17} \,\mathrm{W \, cm^{-2} \, \mu m^2}$ for incident angle $\theta = 45^\circ - 50^\circ$ \cite{PhysRevLett.59.52,PhysRevLett.68.1535,PhysRevLett.73.664}. However, the condition of this experiment is below $I_0 \lambda^2 \leq 10^{17}$ and we expect insignificant vacuum heating (resonance absorption is also negligible given the ultra-steep density profile). Therefore, we expect that there is no need for a PIC simulation and hydrodynamic simulation is sufficient to predict the shock velocities for our intensities. The laser absorption and shock generation can be described in a few stages, depending on the laser intensity profile. Note that the peak of the pulse is located at 360 fs from the origin of the scale.

The rising edge of the extreme contrast SHG laser pulse has intensities around $10^{12} \,\mathrm{W \, cm^{-2}}$, sufficient to generate plasma on the glass surface. The reflectivity of the plasma is estimated to be $R \approx 0.23$ (see SM for the estimation of reflectivity \cite{10.1063/1.1447555}), consistent with the measurement shown in Fig. \ref{fig:reflectivity} between $-5 \, \mathrm{ps} < t < -2 \, \mathrm{ps}$. The absorption mechanism within this time range can be attributed to Fresnel-like absorption or skin heating for the density scale length in the limit $L\rightarrow 0$.

At $t > -2 \, \mathrm{ps}$, the ablation begins and the plasma expands outward with density scale length $L=c_s t$, where $c_s=(\gamma Z k T_e/m_i)^{1/2}$ is the ion acoustic velocity, where $\gamma$ is the heat capacity ratio. The subsequent absorption of the laser light is described by inverse bremsstrahlung. Assuming a fully ionized triatomic gas with $\gamma = 1.33$ and $T_e \geq  2.4 \, \mathrm{eV}$, the plasma expands outward with velocity $\geq 1.2 \times 10^{6} \, \mathrm{cm \, s^{-1}}$. The outward expansion velocity is shown in Fig. \ref{fig:reflectivity} (b) in blue points.

At $t \simeq 144 \, \mathrm{fs}$, the laser intensity raising edge is at the order of $ 10^{16} \, \mathrm{W \, cm^{-2}}$ as shown in Fig. \ref{fig:Y} (a). However, plasma expansion does not produce an ablation pressure that is sufficient to generate a shock. This is evident in Figs. \ref{fig:Y} (b) and (c), where we only observe the compression of the electron density at $t \simeq 144 \, \mathrm{fs}$. The maximum total plasma pressure, as depicted in Fig. \ref{fig:Y} (d), is only approximately $10 \, \mathrm{Mbar}$. We note that the radiation emission pressure from the plasma is negligible at this time.

At $t \simeq 356 \, \mathrm{fs}$, the peak of the pulse reached the target surface. The total pressure at this moment reached $30 \, \mathrm{Mbar}$. At $t \simeq 500 \, \mathrm{fs}$ from the origin the laser pulse is reflected and exits the target surface with a maximum pressure of $50 \, \mathrm{Mbar}$. The pressure begins to decrease at the descending edge ($t \simeq 600 \, \mathrm{fs}$). After this, the laser has completely left the target while the shock continues to propagate into the target. The electron shock front brings the critical density surface inward. This results in the redshift beyond $t \geq 1 \, \mathrm{ps}$ as shown in Fig. \ref{fig:reflectivity} (b) and (c) in red points. The predictions of the velocity of the critical density surface using solely hydrodynamic simulation are consistent with the experimental measurements. This implies that the shock generation is purely hydrodynamic nature.

The shock generation can be attributed to the impulsive excitation by the radiation pressure of an extreme-contrast short-pulse laser on the steep density gradient target. The net radiation pressure on a target can be expressed as $P_\mathrm{rad} = (1 + R) I_0/c $, where $R$ is the reflectivity of the target. The first term represents the pressure exerted by the incident wave, while the second term accounts for the pressure exerted by the reflected wave. This also explains why the plasma pressure peaks during reflection at $t \simeq 500 \, \mathrm{fs}$. The net radiation pressure is $P_\mathrm{rad} = 200 \, \mathrm{Mbar}$ for $I_0 = 3 \times 10^{17} \, \mathrm{W \, cm^{-2}}$ and $R \sim 1$.

The radiation pressure exerts a force $F = P_{\mathrm{rad}} A $ on the skin depth across the focus spot area, $A$. The acceleration is then, $a = F/m = P_{\mathrm{rad}} A/(\rho A  \delta) = P_{\mathrm{rad}}/(\rho \delta)$. The maximum velocity by this impulsive force reaches $v =  P_{\mathrm{rad}} \Delta\tau/(\rho \delta) \approx 2.2 \times 10^7 \, \mathrm{cm \, s^{-1}}$ with fully-ionized plasma skin depth $\delta = c/\omega_{pe} \approx 6.4 \, \mathrm{nm}$ and pulse duration $\Delta \tau = 186 \, \mathrm{fs}$. For $I_0 = 1.3 \times 10^{18} \, \mathrm{W \, cm^{-2}}$, the radiation pressure is $P_\mathrm{rad} = 866 \, \mathrm{Mbar}$.  The impulsive velocity is $v \approx 5.1 \times 10^7 \, \mathrm{cm \, s^{-1}}$ for $\Delta \tau = 100 \, \mathrm{fs}$, and $R \sim 1$.

The shock structure in both the electron and mass density profiles is almost at the same location. This eliminates the possibility of ion motion triggered by the charge separation field in the laser hole-boring (HB) regime \cite{Iwata2018}. The HB regime is characterized by ion motion driven by the charge separation field, which is absent in our scenario. Additionally, the shock structure in our case differs from the ablation-generated shock, where the shock is generated by the ablation pressure, and the ion motion is driven by the ablation front. To the best of our knowledge, this is the first time that the impulsive excitation of a solid by an extreme contrast, high-intensity, non-relativistic femtosecond laser pulse has been studied. A key finding was that the shock generation is purely hydrodynamic, caused by the radiation pressure exerted by the laser main pulse. This contrasts with the ablation-generated shock or hole-boring regime, where the latter primarily involves the charge separation field. This represents a novel mechanism for shock generation in the extreme contrast regime at non-relativistic intensities.

In summary, we have presented the impulsive shock generation on solid density plasma, which is made possible with the extreme contrast of the pulse generated by a high contrast fundamental pulse. This pressure is analogous to the motion of a solid wall under an 'instantly' applied force. The inward motion of the bulk material is present way longer than the duration of the pulse. The compression happens until the energy transferred by the impulse becomes equal to the compressional energy of the bulk high-density plasma. Additionally, the hydrodynamic nature on long picosecond time scale of the problem distinguishes the current scenario from all earlier measurements on shocks.

\section*{Acknowledgments}
GRK acknowledges major support for this research from the grant“Physics and Astronomy (Project Identification No. RTI4002) Department of Atomic Energy,  Tata Institute of Fundamental Research" and partially from the grant JBR/2021/00039  of the Anusandhan  National Research Foundation (ANRF), both of the Government of India. JFO acknowledges EuroHPC Joint Undertaking for awarding us access to Karolina at IT4Innovations (VŠB-TU), Czechia under project number EHPC-REG-2023R02-006 (DD-23-157); Ministry of Education, Youth and Sports of the Czech Republic through the e-INFRA CZ (ID:90140). These works were partly supported by Contract No. PN23210105 funded by the Romanian Ministry of Research, Innovation and Digitalization and of the Extreme Light Infrastructure Nuclear Physics Phase II, a project co-financed by the Romanian Government and the European Union through the European Regional Development Fund and the Competitiveness Operational Program (Grant No. 1/07.07.2016, COP, ID 1334). Additionally, partial support was given by JSPS Core-to-Core Program, Grant Number JPJSCCA20230003.

\bibliography{apssamp}
\end{document}